\documentclass[a4paper,12pt]{article}
\usepackage{a4wide, amsfonts, epsfig}

\newcommand{\omegabf}{\mbox{\boldmath $\omega$}}
\newcommand{\Omegabf}{\mbox{\boldmath $\Omega$}}
\newcommand{\Sigmabf}{\mbox{\boldmath $\Sigma$}}
\newcommand{\sigmabf}{\mbox{\boldmath $\sigma$}}

\begin{document}

\title{\vskip -70pt
  \begin{flushright}
   {\normalsize TCDMATH 05-08, {\tt hep-th/0509098}}
  \end{flushright}
  \vskip 15pt {\bf A zero-mode quantization of the Skyrmion} \author{
{\large Conor Houghton and Shane Magee},\\ {\normalsize {\sl School of
Mathematics, Trinity College Dublin, Dublin 2, Ireland}}\\{\normalsize
{\sl Email: houghton@maths.tcd.ie, magees@maths.tcd.ie}}}
\date{{\large 6 October, 2005}}} \maketitle
\begin{abstract}
In the semi-classical approach to the Skyrme model, nuclei are
approximated by quantum mechanical states on a finite-dimensional
space of field configurations; in zero-mode quantization this space is
generated by rotations and isorotations. Here, simulated annealing is
used to find the axially symmetric Skyrme configuration which
extremizes the zero-mode quantized energy for the nucleon.
\end{abstract}

\section{Introduction}

The Skyrme model is an effective theory of pions and nucleons. It is a
non-linear field theory in which nuclei correspond, classically, to
topological soliton solutions, called Skyrmions.  The model is
non-renormalizable and so the approach usually taken is to reduce to
a finite-dimensional space of Skyrme configurations before
quantizing. In this approach, quantum-mechanical states on a space of
topological charge $B$ Skyrme configurations model baryon number $B$
nuclei at low energies.

The Skyrme Lagrangian is acted on invariantly by rotations and
isorotations; this generates from a single Skyrme configuration a
space of energy-degenerate configurations. This space is known
as the space of zero modes and quantization on this space is called
zero-mode quantization. This approach began with the seminal paper
\cite{ANW:1983} where the nucleon and delta are constructed as quantum 
mechanical states on the zero-mode space generated from the classical 
minimum energy configuration of unit baryon number.

In this paper, the spin-half, isospin-half quantum Hamiltonian is
calculated on the zero-mode space generated from a general axially
symmetric configuration and simulated annealing is used to find the
Skyrme configuration that minimizes the energy of the lowest
state. 

This approach has previously been considered for the zero-mode space
of a general spherical symmetric configuration
\cite{RSWW:1986,BR:1985}. In this space the quantum
Hamiltonian reduces to a scalar, here it is a matrix. Aspects of our
approach are also shared with a very recent paper \cite{BKS:2005}
which appeared when our paper was in preparation. The quantum
Hamiltonian used in that paper is a scalar ansatz motivated by the
spherical Hamiltonian used in \cite{ANW:1983} and differs from the
Hamiltonian we derive here. In fact, we will see that this will not
make a significant difference, the numerical results obtained for the
nucleon in both papers do
not differ much from each other or from what would be calculated using
the classical minimum. However, the approach here is more direct than
the approach described in \cite{BKS:2005} and can be generalized to
higher charge nuclei.

\section{Quantization procedure}

Written in terms of the vector currents $R_{\mu} = \partial_{\mu} U
U^{\dagger}$ of an SU(2) field $U\left(\bf{x}\right)$, the Skyrme
model has the Lagrangian
\begin{equation}
L = \int \mbox{d}^3 {\bf x} \ \left[ \frac{- F_\pi^2}{16} \mbox{Tr} \left(R_\mu R^\mu  \right)
+ \frac{1}{32 e^2} \mbox{Tr} \left(
[ R_\mu, R_\nu][R^\mu, R^\nu]
\right)
+\frac{1}{8} m_\pi^2 F_\pi^2 \mbox{ Tr} \left(U -1\right)\right],
\end{equation}
where $m_{\pi}, F_{\pi}$ and $e$ are parameters that are adjusted to
fit experimental data. Using $F_{\pi}/4 e$ as our unit of energy and $2/e F_{\pi}$ as our unit of length we obtain
\begin{equation}\label{cllagrangian}
L= \int \mbox{d}^3 {\bf x} \ \left[ - \frac{1}{2} \mbox{ Tr} \left(R_\mu R^\mu \right)
+ \frac{1}{16} \mbox{ Tr} \left([ R_\mu, R_\nu][R^\mu, R^\nu]\right)
+\left(\frac{2 m_{\pi} }{ F_{\pi} e}\right)^2  \mbox{ Tr} \left(U -1 \right)\right],
\end{equation}
 The Skyrmion mass for a static field
$U_s\left(\bf{x}\right)$ can be derived from this Lagrangian and is
\begin{equation}\label{clenergy}
M = \int \mbox{ d}^3 {\bf x} \
\left[ -\frac{1}{2} \mbox{ Tr} \left(R_i R_i \right)
    - \frac{1}{16}\left( [ R_i, R_j][R_i, R_j]\right) 
-\left(\frac{2 m_{\pi} }{ F_{\pi} e}\right)^2 \mbox{ Tr} \left(U -1 \right) \right].
\end{equation}

We wish to quantize the rotational and isorotational degrees of
freedom. Rather than acting on a specific Skyrme configuration,
we want to consider the zero-mode space of fields generated from a
general static configuration $U_s\left(\bf{x}\right)$ by isorotation
$C$ and rotation $D$:
\begin{equation}\label{zeromode}
   U \left( \bf{x} \right)
= C U_{s} \left({\bf x}^D \right)  C^{\dagger},
\end{equation}
where $C$ is in the $2 \times 2$ representation,
\begin{equation}
x_i^D=D_{ij}x_j,
\end{equation}
and $D_{i j}$ is a three-dimensional matrix representation of $D$.
Since rotation and isorotation are symmetries of the original
Lagrangian, these configurations are all energy-degenerate. The
effective Lagrangian on this restricted space of configurations can
be calculated by allowing $C$ and $D$ to depend on time, giving $L =
-M + L_{rot}$, where $L_{rot}$ is the kinetic Lagrangian
\begin{equation}\label{clrot}
L_{rot}= \frac{1}{2} \Omega_i U_{ij} \Omega_j
+ \frac{1}{2} \omega_i V_{ij} \omega_j - \omega_i W_{ij} \Omega_j,
\end{equation}
with the rotational and isorotational angular velocities $\omegabf$ and $\Omegabf$ given by 
\begin{eqnarray}
   \Omega_i&=&-i\rm{Tr}\left(\sigma_i C^{\dagger}\dot{C}\right), \cr
   \omega_i&=&-i\rm{Tr}\left(\sigma_i D^{\dagger}\dot{D}\right), 
\end{eqnarray}
and the moment of inertia tensors $U_{ij}$, $V_{ij}$ and $W_{ij}$ given by the following integrals 
\begin{eqnarray}\label{clinertia}
U_{ij} &=& -\int \mbox{ d}^3 {\bf x}\
\left[\mbox{ Tr} \left(T_i T_j \right) +
\frac{1}{4} \mbox{ Tr} \left([R_k,T_i][R_k,T_j] \right)
\right],\cr 
V_{ij}&=& -\epsilon_{ilm}\epsilon_{jpq}\int \mbox{ d}^3 {\bf x}\ x_l x_p
\left[\mbox{ Tr} \left(R_m R_q \right)
+\frac{1}{4} \mbox{ Tr} \left([R_k,R_m][R_k,R_q]\right)
\right],\cr
W_{ij} &=& \epsilon_{jlm} \int \mbox{ d}^3 {\bf x}\ x_l\left[
\mbox{ Tr} \left(T_i R_m \right)
+\frac{1}{4} \mbox{ Tr} \left([R_k,T_i][R_k,R_m]\right)
\right].
\end{eqnarray}
and 
\begin{equation}
 T_i=i\left[\frac{\sigma_i}{2},U\right]U^{\dagger},
\end{equation}
where the $\sigma_i$ are the usual Pauli matrices.

In this paper, we will restrict our discussion to axially symmetric
Skyrmion solutions. Numerical simulations indicate that there is also
a reflection symmetry in the $xy$-plane and so the principal axes of
inertia can be taken as the standard orthogonal axes, with
$U_{ij}=V_{ij}=W_{ij}=0$ where $i \neq j$, and we can set
$U_{ii}=U_{i}$ and so forth.  By inspection, cylindrical symmetry in
the $xy$-plane will also mean
\begin{equation}\label{reln1}
U_1=U_2,\quad  V_1=V_2,\quad W_1=W_2.
\end{equation}
We can use axial symmetry to establish an additional identification between the normal moments of inertia (see Appendix):
\begin{equation}\label{reln2}
U_3=V_3=W_3
\end{equation}
Applying these restrictions to $L_{rot}$ (\ref{clrot}) we get
\begin{equation}
L_{rot} = \frac{1}{2} \left( \Omega_1^2 + \Omega_2^2 \right) U_2 +
\frac{1}{2} \left( \omega_1^2 + \omega_2^2 \right) V_2 + \frac{1}{2} \left( \Omega_3 - \omega_3 \right)^2 U_3 - 
 \left( \Omega_1 \omega_1 + \Omega_2 \omega_2 \right) W_2 
\end{equation}
or, written as a sum of complete squares,
\begin{eqnarray}\label{squareeqn}
L_{rot}&=& \frac{1}{2} \left( V_2 -\frac{W_2^2}{U_2} \right)\left(
\omega_1^2 + \omega_2^2 \right) + \frac{1}{2}\left( \Omega_3 -
\omega_3 \right)^2 U_3 \nonumber\\[5pt]&& +\frac{1}{2}\left[ \left(
\Omega_1 -\frac{W_2}{U_2}\omega_1 \right)^2 +\left( \Omega_2
-\frac{W_2}{U_2}\omega_2 \right) ^2\right]U_2.
\end{eqnarray}

The rotation and isorotation angular momentum vectors {\bf L} and {\bf
K} canonically conjugate to $\omegabf$ and $\Omegabf$ are
\begin{eqnarray}\label{LandK}
{\bf L}&=& \frac{\partial L_{rot}}{\partial \omegabf }
 =\left(
 V_2 \omega_1-W_2 \Omega_1 ,
 V_2 \omega_2-W_2 \Omega_2 ,
 U_3 \left( \omega_3 - \Omega_3 \right)\right),\nonumber \\[5pt]
{\bf K}&=& \frac{\partial L_{rot}}{\partial \Omegabf } 
=\left(
 U_2 \Omega_1-W_2 \omega_1 ,
 U_2 \Omega_2-W_2 \omega_2 ,
 -U_3 \left( \omega_3 - \Omega_3 \right)\right). 
\end{eqnarray}
Note that $L_3$=$-K_3$: upon quantization, this is the axially
symmetric condition expressed in operator form. Our approach will be
to find the minimum energy Skyrmion such that its energy eigenstate is
also a null eigenstate of $L_3 + K_3$.  From (\ref{squareeqn}) and
(\ref{LandK}), the Hamiltonian for the rotational and isorotational
degrees of freedom can now be calculated:
\begin{eqnarray}\label{hamiltonian}
H &=& \mathbf{L}.\omegabf +\mathbf{K}.\Omegabf - L_{rot} \cr&&\cr
 &=&\frac{1}{2} \left[
\frac{\left( L_1+\frac{W_2}{U_2} K_1 \right)^2}{V_2-\frac{W_2^2}{U_2}}+
\frac{\left( L_2+\frac{W_2}{U_2} K_2 \right) ^2}{V_2-\frac{W_2^2}{U_2}}+
\frac{K_1^2}{U_1} +\frac{K_2^2}{U_2}+\frac{L_3^2}{U_3}
\right].
\end{eqnarray}

For a spin-$n$,isospin-$n$ particle, the angular momentum operators
$\mathbf{L}$ can be written as $\left( 2n+1 \right) \times \left( 2n+1
\right)$ dimensional matrix representation $\Sigma_1^L, \Sigma_2^L,
\Sigma_3^L$ of SU(2), as can the isospin operators $\Sigma_1^K,
\Sigma_2^K, \Sigma_3^K $ of ${\bf K}$ for an isospin-$n$ particle. If
we define our quantum state using the $\vert l,l_3\rangle\otimes\vert
k,k_3\rangle$ basis, $l$, $l_3$ and $k$, $k_3$ being the quantum
numbers for $\mathbf{L}$ and $\mathbf{K}$ respectively, we can
embed $\mathbf{L}$ and $\mathbf{K}$ into the resulting SO(3)$^L
\times$ SO(3)$^K$ direct product space:
\begin{eqnarray}\label{matrixLK}
\mathbf{L}&\mapsto&\hbar \Sigmabf^L \otimes I_{2n+1}  \nonumber \\[5pt]
\mathbf{K}&\mapsto&I_{2n+1} \otimes \hbar \Sigmabf^K   
\end{eqnarray}
where $I_{2n+1}$ is the ($2n+1$)-dimensional identity matrix.

We can now find the lowest energy nucleon state; first we insert the
spin-half, isospin-half matrix representation of ${\bf L}$ and ${\bf
K}$ into the Hamiltonian (\ref{hamiltonian}) to get:
\begin{equation}
H=\frac{\hbar^2}{4}\left(
\begin{array}{cccc} \kappa_1 & 0 & 0 & 0 \\ 
               0 & \kappa_1 & \kappa_2 & 0 \\
                0 & \kappa_2 & \kappa_1 & 0 \\ 
                       0 & 0 & 0 & \kappa_1 \end{array}
\right)
\end{equation}
where 
\begin{equation}
\kappa_1 =
\frac{1+\left(\frac{W_2}{U_2}\right)^2}{V_2-\frac{W_2^2}{U_2}
}+\frac{1}{U_2}+\frac{1}{2 U_3}
\end{equation}
and
\begin{equation}
 \kappa_2 =\frac{2\frac{W_2}{U_2}}{V_2-\frac{W_2^2}{U_2}}.  
\end{equation}

There are two eigenvectors of $H$ which are also eigenvectors of
$L_3+K_3$ with eigenvalue zero:
 \begin{equation}
|0,0\rangle \equiv
\frac{1}{\sqrt{2}}\left(|\frac{1}{2},-\frac{1}{2}\rangle \otimes
|\frac{1}{2},\frac{1}{2}\rangle- |\frac{1}{2},\frac{1}{2}\rangle
\otimes
|\frac{1}{2},-\frac{1}{2}\rangle\right),\quad E_{0,0}=\frac{\hbar^2}{4}\left(\kappa_1-\kappa_2\right)
\end{equation}
and
\begin{equation}
|1,0\rangle \equiv
\frac{1}{\sqrt{2}}\left(|\frac{1}{2},-\frac{1}{2}\rangle \otimes
|\frac{1}{2},\frac{1}{2}\rangle+ |\frac{1}{2},\frac{1}{2}\rangle
\otimes |\frac{1}{2},-\frac{1}{2}\rangle\right),
\quad E_{1,0}=\frac{\hbar^2}{4}\left(\kappa_1+\kappa_2\right) 
\end{equation}
The first eigenvector is a spherically symmetric state and also has
the lower energy since $\kappa_1$ and $\kappa_2$ are always positive; hence
\begin{equation}\label{lowestE}
E_N= \frac{\hbar^2}{4}\left[
        \frac{\left(1-\frac{W_2}{U_2}\right)^2}{V_2-\frac{W_2^2}{U_2}
        }+\frac{1}{U_2} +\frac{1}{2 U_3}\right]
\end{equation}
and is therefore the energy of an axially symmetric nucleon. We see
that in the spherically symmetric case $U_2=U_3=W_2=V_2$,
and $E_N$ reduces to the rotational energy formula obtained in
\cite{ANW:1983}:
\begin{equation}
E_N^{sym} = \frac{\hbar^2}{2
U_3}l\left(l+1\right)=\frac{3}{4}\frac{\hbar}{2 \Lambda}
\end{equation}
where $\Lambda=\frac{1}{3}\left(U_1+U_2+U_3\right)=U_3$. The energies
of all other eigenstates go to infinity in the spherically symmetric limit.

The energy, $M + E_N$, of the quantum state would be difficult to
extremize using gradient-based methods; instead, simulated annealing \cite{HSW:2000}
is used to find the Skyrmion configuration that minimizes this
energy. Since the configuration is assumed to be axially symmetric,
the cartoon method \cite{ABBHSTT:2001} is used. 
The configuration is annealed on a
quarter-disk two-dimensional lattice with a radius of 250 lattice
points and a lattice spacing of 0.06. A variant of the Adaptive
Simulated Annealing probability distribution \cite{In:1996} is used
for the field perturbations; this seems to improve the speed of
convergence and allows an exponential cooling schedule. The algorithm
needs an initial configuration to perturb; any configuration of unit
baryon number will suffice, and the ansatz given in \cite{S:1992} is
probably the easiest to implement.

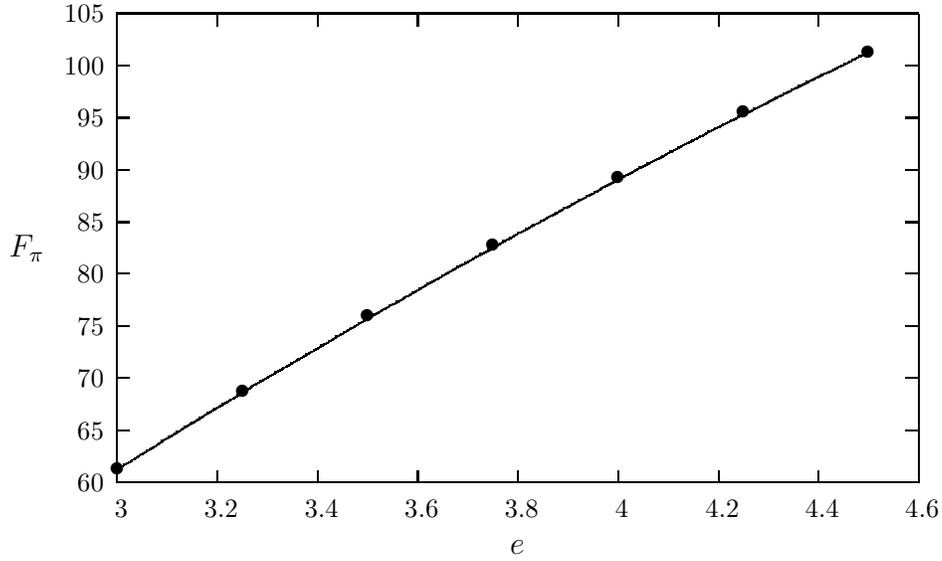
\begin{figure}
\begin{center}
% GNUPLOT: LaTeX picture
\setlength{\unitlength}{0.240900pt}
\ifx\plotpoint\undefined\newsavebox{\plotpoint}\fi
\sbox{\plotpoint}{\rule[-0.200pt]{0.400pt}{0.400pt}}%
\begin{picture}(1500,900)(0,0)
\font\gnuplot=cmr10 at 10pt
\gnuplot
\sbox{\plotpoint}{\rule[-0.200pt]{0.400pt}{0.400pt}}%
\put(181.0,123.0){\rule[-0.200pt]{4.818pt}{0.400pt}}
\put(161,123){\makebox(0,0)[r]{ 60}}
\put(1419.0,123.0){\rule[-0.200pt]{4.818pt}{0.400pt}}
\put(181.0,205.0){\rule[-0.200pt]{4.818pt}{0.400pt}}
\put(161,205){\makebox(0,0)[r]{ 65}}
\put(1419.0,205.0){\rule[-0.200pt]{4.818pt}{0.400pt}}
\put(181.0,287.0){\rule[-0.200pt]{4.818pt}{0.400pt}}
\put(161,287){\makebox(0,0)[r]{ 70}}
\put(1419.0,287.0){\rule[-0.200pt]{4.818pt}{0.400pt}}
\put(181.0,369.0){\rule[-0.200pt]{4.818pt}{0.400pt}}
\put(161,369){\makebox(0,0)[r]{ 75}}
\put(1419.0,369.0){\rule[-0.200pt]{4.818pt}{0.400pt}}
\put(181.0,451.0){\rule[-0.200pt]{4.818pt}{0.400pt}}
\put(161,451){\makebox(0,0)[r]{ 80}}
\put(1419.0,451.0){\rule[-0.200pt]{4.818pt}{0.400pt}}
\put(181.0,532.0){\rule[-0.200pt]{4.818pt}{0.400pt}}
\put(161,532){\makebox(0,0)[r]{ 85}}
\put(1419.0,532.0){\rule[-0.200pt]{4.818pt}{0.400pt}}
\put(181.0,614.0){\rule[-0.200pt]{4.818pt}{0.400pt}}
\put(161,614){\makebox(0,0)[r]{ 90}}
\put(1419.0,614.0){\rule[-0.200pt]{4.818pt}{0.400pt}}
\put(181.0,696.0){\rule[-0.200pt]{4.818pt}{0.400pt}}
\put(161,696){\makebox(0,0)[r]{ 95}}
\put(1419.0,696.0){\rule[-0.200pt]{4.818pt}{0.400pt}}
\put(181.0,778.0){\rule[-0.200pt]{4.818pt}{0.400pt}}
\put(161,778){\makebox(0,0)[r]{ 100}}
\put(1419.0,778.0){\rule[-0.200pt]{4.818pt}{0.400pt}}
\put(181.0,860.0){\rule[-0.200pt]{4.818pt}{0.400pt}}
\put(161,860){\makebox(0,0)[r]{ 105}}
\put(1419.0,860.0){\rule[-0.200pt]{4.818pt}{0.400pt}}
\put(181.0,123.0){\rule[-0.200pt]{0.400pt}{4.818pt}}
\put(181,82){\makebox(0,0){ 3}}
\put(181.0,840.0){\rule[-0.200pt]{0.400pt}{4.818pt}}
\put(338.0,123.0){\rule[-0.200pt]{0.400pt}{4.818pt}}
\put(338,82){\makebox(0,0){ 3.2}}
\put(338.0,840.0){\rule[-0.200pt]{0.400pt}{4.818pt}}
\put(496.0,123.0){\rule[-0.200pt]{0.400pt}{4.818pt}}
\put(496,82){\makebox(0,0){ 3.4}}
\put(496.0,840.0){\rule[-0.200pt]{0.400pt}{4.818pt}}
\put(653.0,123.0){\rule[-0.200pt]{0.400pt}{4.818pt}}
\put(653,82){\makebox(0,0){ 3.6}}
\put(653.0,840.0){\rule[-0.200pt]{0.400pt}{4.818pt}}
\put(810.0,123.0){\rule[-0.200pt]{0.400pt}{4.818pt}}
\put(810,82){\makebox(0,0){ 3.8}}
\put(810.0,840.0){\rule[-0.200pt]{0.400pt}{4.818pt}}
\put(967.0,123.0){\rule[-0.200pt]{0.400pt}{4.818pt}}
\put(967,82){\makebox(0,0){ 4}}
\put(967.0,840.0){\rule[-0.200pt]{0.400pt}{4.818pt}}
\put(1125.0,123.0){\rule[-0.200pt]{0.400pt}{4.818pt}}
\put(1125,82){\makebox(0,0){ 4.2}}
\put(1125.0,840.0){\rule[-0.200pt]{0.400pt}{4.818pt}}
\put(1282.0,123.0){\rule[-0.200pt]{0.400pt}{4.818pt}}
\put(1282,82){\makebox(0,0){ 4.4}}
\put(1282.0,840.0){\rule[-0.200pt]{0.400pt}{4.818pt}}
\put(1439.0,123.0){\rule[-0.200pt]{0.400pt}{4.818pt}}
\put(1439,82){\makebox(0,0){ 4.6}}
\put(1439.0,840.0){\rule[-0.200pt]{0.400pt}{4.818pt}}
\put(181.0,123.0){\rule[-0.200pt]{303.052pt}{0.400pt}}
\put(1439.0,123.0){\rule[-0.200pt]{0.400pt}{177.543pt}}
\put(181.0,860.0){\rule[-0.200pt]{303.052pt}{0.400pt}}
\put(40,491){\makebox(0,0){$F_{\pi}$}}
\put(810,21){\makebox(0,0){$e$}}
\put(181.0,123.0){\rule[-0.200pt]{0.400pt}{177.543pt}}
\put(181,144){\usebox{\plotpoint}}
\multiput(181.00,144.59)(0.874,0.485){11}{\rule{0.786pt}{0.117pt}}
\multiput(181.00,143.17)(10.369,7.000){2}{\rule{0.393pt}{0.400pt}}
\multiput(193.00,151.59)(0.874,0.485){11}{\rule{0.786pt}{0.117pt}}
\multiput(193.00,150.17)(10.369,7.000){2}{\rule{0.393pt}{0.400pt}}
\multiput(205.00,158.59)(0.758,0.488){13}{\rule{0.700pt}{0.117pt}}
\multiput(205.00,157.17)(10.547,8.000){2}{\rule{0.350pt}{0.400pt}}
\multiput(217.00,166.59)(0.874,0.485){11}{\rule{0.786pt}{0.117pt}}
\multiput(217.00,165.17)(10.369,7.000){2}{\rule{0.393pt}{0.400pt}}
\multiput(229.00,173.59)(0.874,0.485){11}{\rule{0.786pt}{0.117pt}}
\multiput(229.00,172.17)(10.369,7.000){2}{\rule{0.393pt}{0.400pt}}
\multiput(241.00,180.59)(0.692,0.488){13}{\rule{0.650pt}{0.117pt}}
\multiput(241.00,179.17)(9.651,8.000){2}{\rule{0.325pt}{0.400pt}}
\multiput(252.00,188.59)(0.874,0.485){11}{\rule{0.786pt}{0.117pt}}
\multiput(252.00,187.17)(10.369,7.000){2}{\rule{0.393pt}{0.400pt}}
\multiput(264.00,195.59)(0.874,0.485){11}{\rule{0.786pt}{0.117pt}}
\multiput(264.00,194.17)(10.369,7.000){2}{\rule{0.393pt}{0.400pt}}
\multiput(276.00,202.59)(0.758,0.488){13}{\rule{0.700pt}{0.117pt}}
\multiput(276.00,201.17)(10.547,8.000){2}{\rule{0.350pt}{0.400pt}}
\multiput(288.00,210.59)(0.874,0.485){11}{\rule{0.786pt}{0.117pt}}
\multiput(288.00,209.17)(10.369,7.000){2}{\rule{0.393pt}{0.400pt}}
\multiput(300.00,217.59)(0.874,0.485){11}{\rule{0.786pt}{0.117pt}}
\multiput(300.00,216.17)(10.369,7.000){2}{\rule{0.393pt}{0.400pt}}
\multiput(312.00,224.59)(0.758,0.488){13}{\rule{0.700pt}{0.117pt}}
\multiput(312.00,223.17)(10.547,8.000){2}{\rule{0.350pt}{0.400pt}}
\multiput(324.00,232.59)(0.874,0.485){11}{\rule{0.786pt}{0.117pt}}
\multiput(324.00,231.17)(10.369,7.000){2}{\rule{0.393pt}{0.400pt}}
\multiput(336.00,239.59)(0.874,0.485){11}{\rule{0.786pt}{0.117pt}}
\multiput(336.00,238.17)(10.369,7.000){2}{\rule{0.393pt}{0.400pt}}
\multiput(348.00,246.59)(0.874,0.485){11}{\rule{0.786pt}{0.117pt}}
\multiput(348.00,245.17)(10.369,7.000){2}{\rule{0.393pt}{0.400pt}}
\multiput(360.00,253.59)(0.874,0.485){11}{\rule{0.786pt}{0.117pt}}
\multiput(360.00,252.17)(10.369,7.000){2}{\rule{0.393pt}{0.400pt}}
\multiput(372.00,260.59)(0.758,0.488){13}{\rule{0.700pt}{0.117pt}}
\multiput(372.00,259.17)(10.547,8.000){2}{\rule{0.350pt}{0.400pt}}
\multiput(384.00,268.59)(0.798,0.485){11}{\rule{0.729pt}{0.117pt}}
\multiput(384.00,267.17)(9.488,7.000){2}{\rule{0.364pt}{0.400pt}}
\multiput(395.00,275.59)(0.874,0.485){11}{\rule{0.786pt}{0.117pt}}
\multiput(395.00,274.17)(10.369,7.000){2}{\rule{0.393pt}{0.400pt}}
\multiput(407.00,282.59)(0.874,0.485){11}{\rule{0.786pt}{0.117pt}}
\multiput(407.00,281.17)(10.369,7.000){2}{\rule{0.393pt}{0.400pt}}
\multiput(419.00,289.59)(0.874,0.485){11}{\rule{0.786pt}{0.117pt}}
\multiput(419.00,288.17)(10.369,7.000){2}{\rule{0.393pt}{0.400pt}}
\multiput(431.00,296.59)(0.874,0.485){11}{\rule{0.786pt}{0.117pt}}
\multiput(431.00,295.17)(10.369,7.000){2}{\rule{0.393pt}{0.400pt}}
\multiput(443.00,303.59)(0.874,0.485){11}{\rule{0.786pt}{0.117pt}}
\multiput(443.00,302.17)(10.369,7.000){2}{\rule{0.393pt}{0.400pt}}
\multiput(455.00,310.59)(0.874,0.485){11}{\rule{0.786pt}{0.117pt}}
\multiput(455.00,309.17)(10.369,7.000){2}{\rule{0.393pt}{0.400pt}}
\multiput(467.00,317.59)(0.874,0.485){11}{\rule{0.786pt}{0.117pt}}
\multiput(467.00,316.17)(10.369,7.000){2}{\rule{0.393pt}{0.400pt}}
\multiput(479.00,324.59)(0.874,0.485){11}{\rule{0.786pt}{0.117pt}}
\multiput(479.00,323.17)(10.369,7.000){2}{\rule{0.393pt}{0.400pt}}
\multiput(491.00,331.59)(0.874,0.485){11}{\rule{0.786pt}{0.117pt}}
\multiput(491.00,330.17)(10.369,7.000){2}{\rule{0.393pt}{0.400pt}}
\multiput(503.00,338.59)(0.874,0.485){11}{\rule{0.786pt}{0.117pt}}
\multiput(503.00,337.17)(10.369,7.000){2}{\rule{0.393pt}{0.400pt}}
\multiput(515.00,345.59)(0.798,0.485){11}{\rule{0.729pt}{0.117pt}}
\multiput(515.00,344.17)(9.488,7.000){2}{\rule{0.364pt}{0.400pt}}
\multiput(526.00,352.59)(0.874,0.485){11}{\rule{0.786pt}{0.117pt}}
\multiput(526.00,351.17)(10.369,7.000){2}{\rule{0.393pt}{0.400pt}}
\multiput(538.00,359.59)(0.874,0.485){11}{\rule{0.786pt}{0.117pt}}
\multiput(538.00,358.17)(10.369,7.000){2}{\rule{0.393pt}{0.400pt}}
\multiput(550.00,366.59)(0.874,0.485){11}{\rule{0.786pt}{0.117pt}}
\multiput(550.00,365.17)(10.369,7.000){2}{\rule{0.393pt}{0.400pt}}
\multiput(562.00,373.59)(0.874,0.485){11}{\rule{0.786pt}{0.117pt}}
\multiput(562.00,372.17)(10.369,7.000){2}{\rule{0.393pt}{0.400pt}}
\multiput(574.00,380.59)(0.874,0.485){11}{\rule{0.786pt}{0.117pt}}
\multiput(574.00,379.17)(10.369,7.000){2}{\rule{0.393pt}{0.400pt}}
\multiput(586.00,387.59)(0.874,0.485){11}{\rule{0.786pt}{0.117pt}}
\multiput(586.00,386.17)(10.369,7.000){2}{\rule{0.393pt}{0.400pt}}
\multiput(598.00,394.59)(0.874,0.485){11}{\rule{0.786pt}{0.117pt}}
\multiput(598.00,393.17)(10.369,7.000){2}{\rule{0.393pt}{0.400pt}}
\multiput(610.00,401.59)(1.033,0.482){9}{\rule{0.900pt}{0.116pt}}
\multiput(610.00,400.17)(10.132,6.000){2}{\rule{0.450pt}{0.400pt}}
\multiput(622.00,407.59)(0.874,0.485){11}{\rule{0.786pt}{0.117pt}}
\multiput(622.00,406.17)(10.369,7.000){2}{\rule{0.393pt}{0.400pt}}
\multiput(634.00,414.59)(0.874,0.485){11}{\rule{0.786pt}{0.117pt}}
\multiput(634.00,413.17)(10.369,7.000){2}{\rule{0.393pt}{0.400pt}}
\multiput(646.00,421.59)(0.874,0.485){11}{\rule{0.786pt}{0.117pt}}
\multiput(646.00,420.17)(10.369,7.000){2}{\rule{0.393pt}{0.400pt}}
\multiput(658.00,428.59)(0.798,0.485){11}{\rule{0.729pt}{0.117pt}}
\multiput(658.00,427.17)(9.488,7.000){2}{\rule{0.364pt}{0.400pt}}
\multiput(669.00,435.59)(1.033,0.482){9}{\rule{0.900pt}{0.116pt}}
\multiput(669.00,434.17)(10.132,6.000){2}{\rule{0.450pt}{0.400pt}}
\multiput(681.00,441.59)(0.874,0.485){11}{\rule{0.786pt}{0.117pt}}
\multiput(681.00,440.17)(10.369,7.000){2}{\rule{0.393pt}{0.400pt}}
\multiput(693.00,448.59)(0.874,0.485){11}{\rule{0.786pt}{0.117pt}}
\multiput(693.00,447.17)(10.369,7.000){2}{\rule{0.393pt}{0.400pt}}
\multiput(705.00,455.59)(0.874,0.485){11}{\rule{0.786pt}{0.117pt}}
\multiput(705.00,454.17)(10.369,7.000){2}{\rule{0.393pt}{0.400pt}}
\multiput(717.00,462.59)(1.033,0.482){9}{\rule{0.900pt}{0.116pt}}
\multiput(717.00,461.17)(10.132,6.000){2}{\rule{0.450pt}{0.400pt}}
\multiput(729.00,468.59)(0.874,0.485){11}{\rule{0.786pt}{0.117pt}}
\multiput(729.00,467.17)(10.369,7.000){2}{\rule{0.393pt}{0.400pt}}
\multiput(741.00,475.59)(0.874,0.485){11}{\rule{0.786pt}{0.117pt}}
\multiput(741.00,474.17)(10.369,7.000){2}{\rule{0.393pt}{0.400pt}}
\multiput(753.00,482.59)(1.033,0.482){9}{\rule{0.900pt}{0.116pt}}
\multiput(753.00,481.17)(10.132,6.000){2}{\rule{0.450pt}{0.400pt}}
\multiput(765.00,488.59)(0.874,0.485){11}{\rule{0.786pt}{0.117pt}}
\multiput(765.00,487.17)(10.369,7.000){2}{\rule{0.393pt}{0.400pt}}
\multiput(777.00,495.59)(0.874,0.485){11}{\rule{0.786pt}{0.117pt}}
\multiput(777.00,494.17)(10.369,7.000){2}{\rule{0.393pt}{0.400pt}}
\multiput(789.00,502.59)(0.943,0.482){9}{\rule{0.833pt}{0.116pt}}
\multiput(789.00,501.17)(9.270,6.000){2}{\rule{0.417pt}{0.400pt}}
\multiput(800.00,508.59)(0.874,0.485){11}{\rule{0.786pt}{0.117pt}}
\multiput(800.00,507.17)(10.369,7.000){2}{\rule{0.393pt}{0.400pt}}
\multiput(812.00,515.59)(1.033,0.482){9}{\rule{0.900pt}{0.116pt}}
\multiput(812.00,514.17)(10.132,6.000){2}{\rule{0.450pt}{0.400pt}}
\multiput(824.00,521.59)(0.874,0.485){11}{\rule{0.786pt}{0.117pt}}
\multiput(824.00,520.17)(10.369,7.000){2}{\rule{0.393pt}{0.400pt}}
\multiput(836.00,528.59)(1.033,0.482){9}{\rule{0.900pt}{0.116pt}}
\multiput(836.00,527.17)(10.132,6.000){2}{\rule{0.450pt}{0.400pt}}
\multiput(848.00,534.59)(0.874,0.485){11}{\rule{0.786pt}{0.117pt}}
\multiput(848.00,533.17)(10.369,7.000){2}{\rule{0.393pt}{0.400pt}}
\multiput(860.00,541.59)(1.033,0.482){9}{\rule{0.900pt}{0.116pt}}
\multiput(860.00,540.17)(10.132,6.000){2}{\rule{0.450pt}{0.400pt}}
\multiput(872.00,547.59)(0.874,0.485){11}{\rule{0.786pt}{0.117pt}}
\multiput(872.00,546.17)(10.369,7.000){2}{\rule{0.393pt}{0.400pt}}
\multiput(884.00,554.59)(1.033,0.482){9}{\rule{0.900pt}{0.116pt}}
\multiput(884.00,553.17)(10.132,6.000){2}{\rule{0.450pt}{0.400pt}}
\multiput(896.00,560.59)(0.874,0.485){11}{\rule{0.786pt}{0.117pt}}
\multiput(896.00,559.17)(10.369,7.000){2}{\rule{0.393pt}{0.400pt}}
\multiput(908.00,567.59)(1.033,0.482){9}{\rule{0.900pt}{0.116pt}}
\multiput(908.00,566.17)(10.132,6.000){2}{\rule{0.450pt}{0.400pt}}
\multiput(920.00,573.59)(1.033,0.482){9}{\rule{0.900pt}{0.116pt}}
\multiput(920.00,572.17)(10.132,6.000){2}{\rule{0.450pt}{0.400pt}}
\multiput(932.00,579.59)(0.798,0.485){11}{\rule{0.729pt}{0.117pt}}
\multiput(932.00,578.17)(9.488,7.000){2}{\rule{0.364pt}{0.400pt}}
\multiput(943.00,586.59)(1.033,0.482){9}{\rule{0.900pt}{0.116pt}}
\multiput(943.00,585.17)(10.132,6.000){2}{\rule{0.450pt}{0.400pt}}
\multiput(955.00,592.59)(0.874,0.485){11}{\rule{0.786pt}{0.117pt}}
\multiput(955.00,591.17)(10.369,7.000){2}{\rule{0.393pt}{0.400pt}}
\multiput(967.00,599.59)(1.033,0.482){9}{\rule{0.900pt}{0.116pt}}
\multiput(967.00,598.17)(10.132,6.000){2}{\rule{0.450pt}{0.400pt}}
\multiput(979.00,605.59)(1.033,0.482){9}{\rule{0.900pt}{0.116pt}}
\multiput(979.00,604.17)(10.132,6.000){2}{\rule{0.450pt}{0.400pt}}
\multiput(991.00,611.59)(0.874,0.485){11}{\rule{0.786pt}{0.117pt}}
\multiput(991.00,610.17)(10.369,7.000){2}{\rule{0.393pt}{0.400pt}}
\multiput(1003.00,618.59)(1.033,0.482){9}{\rule{0.900pt}{0.116pt}}
\multiput(1003.00,617.17)(10.132,6.000){2}{\rule{0.450pt}{0.400pt}}
\multiput(1015.00,624.59)(1.033,0.482){9}{\rule{0.900pt}{0.116pt}}
\multiput(1015.00,623.17)(10.132,6.000){2}{\rule{0.450pt}{0.400pt}}
\multiput(1027.00,630.59)(1.033,0.482){9}{\rule{0.900pt}{0.116pt}}
\multiput(1027.00,629.17)(10.132,6.000){2}{\rule{0.450pt}{0.400pt}}
\multiput(1039.00,636.59)(0.874,0.485){11}{\rule{0.786pt}{0.117pt}}
\multiput(1039.00,635.17)(10.369,7.000){2}{\rule{0.393pt}{0.400pt}}
\multiput(1051.00,643.59)(1.033,0.482){9}{\rule{0.900pt}{0.116pt}}
\multiput(1051.00,642.17)(10.132,6.000){2}{\rule{0.450pt}{0.400pt}}
\multiput(1063.00,649.59)(0.943,0.482){9}{\rule{0.833pt}{0.116pt}}
\multiput(1063.00,648.17)(9.270,6.000){2}{\rule{0.417pt}{0.400pt}}
\multiput(1074.00,655.59)(1.033,0.482){9}{\rule{0.900pt}{0.116pt}}
\multiput(1074.00,654.17)(10.132,6.000){2}{\rule{0.450pt}{0.400pt}}
\multiput(1086.00,661.59)(1.033,0.482){9}{\rule{0.900pt}{0.116pt}}
\multiput(1086.00,660.17)(10.132,6.000){2}{\rule{0.450pt}{0.400pt}}
\multiput(1098.00,667.59)(0.874,0.485){11}{\rule{0.786pt}{0.117pt}}
\multiput(1098.00,666.17)(10.369,7.000){2}{\rule{0.393pt}{0.400pt}}
\multiput(1110.00,674.59)(1.033,0.482){9}{\rule{0.900pt}{0.116pt}}
\multiput(1110.00,673.17)(10.132,6.000){2}{\rule{0.450pt}{0.400pt}}
\multiput(1122.00,680.59)(1.033,0.482){9}{\rule{0.900pt}{0.116pt}}
\multiput(1122.00,679.17)(10.132,6.000){2}{\rule{0.450pt}{0.400pt}}
\multiput(1134.00,686.59)(1.033,0.482){9}{\rule{0.900pt}{0.116pt}}
\multiput(1134.00,685.17)(10.132,6.000){2}{\rule{0.450pt}{0.400pt}}
\multiput(1146.00,692.59)(1.033,0.482){9}{\rule{0.900pt}{0.116pt}}
\multiput(1146.00,691.17)(10.132,6.000){2}{\rule{0.450pt}{0.400pt}}
\multiput(1158.00,698.59)(1.033,0.482){9}{\rule{0.900pt}{0.116pt}}
\multiput(1158.00,697.17)(10.132,6.000){2}{\rule{0.450pt}{0.400pt}}
\multiput(1170.00,704.59)(1.033,0.482){9}{\rule{0.900pt}{0.116pt}}
\multiput(1170.00,703.17)(10.132,6.000){2}{\rule{0.450pt}{0.400pt}}
\multiput(1182.00,710.59)(1.033,0.482){9}{\rule{0.900pt}{0.116pt}}
\multiput(1182.00,709.17)(10.132,6.000){2}{\rule{0.450pt}{0.400pt}}
\multiput(1194.00,716.59)(1.033,0.482){9}{\rule{0.900pt}{0.116pt}}
\multiput(1194.00,715.17)(10.132,6.000){2}{\rule{0.450pt}{0.400pt}}
\multiput(1206.00,722.59)(0.943,0.482){9}{\rule{0.833pt}{0.116pt}}
\multiput(1206.00,721.17)(9.270,6.000){2}{\rule{0.417pt}{0.400pt}}
\multiput(1217.00,728.59)(1.033,0.482){9}{\rule{0.900pt}{0.116pt}}
\multiput(1217.00,727.17)(10.132,6.000){2}{\rule{0.450pt}{0.400pt}}
\multiput(1229.00,734.59)(1.033,0.482){9}{\rule{0.900pt}{0.116pt}}
\multiput(1229.00,733.17)(10.132,6.000){2}{\rule{0.450pt}{0.400pt}}
\multiput(1241.00,740.59)(1.033,0.482){9}{\rule{0.900pt}{0.116pt}}
\multiput(1241.00,739.17)(10.132,6.000){2}{\rule{0.450pt}{0.400pt}}
\multiput(1253.00,746.59)(1.033,0.482){9}{\rule{0.900pt}{0.116pt}}
\multiput(1253.00,745.17)(10.132,6.000){2}{\rule{0.450pt}{0.400pt}}
\multiput(1265.00,752.59)(1.033,0.482){9}{\rule{0.900pt}{0.116pt}}
\multiput(1265.00,751.17)(10.132,6.000){2}{\rule{0.450pt}{0.400pt}}
\multiput(1277.00,758.59)(1.033,0.482){9}{\rule{0.900pt}{0.116pt}}
\multiput(1277.00,757.17)(10.132,6.000){2}{\rule{0.450pt}{0.400pt}}
\multiput(1289.00,764.59)(1.267,0.477){7}{\rule{1.060pt}{0.115pt}}
\multiput(1289.00,763.17)(9.800,5.000){2}{\rule{0.530pt}{0.400pt}}
\multiput(1301.00,769.59)(1.033,0.482){9}{\rule{0.900pt}{0.116pt}}
\multiput(1301.00,768.17)(10.132,6.000){2}{\rule{0.450pt}{0.400pt}}
\multiput(1313.00,775.59)(1.033,0.482){9}{\rule{0.900pt}{0.116pt}}
\multiput(1313.00,774.17)(10.132,6.000){2}{\rule{0.450pt}{0.400pt}}
\multiput(1325.00,781.59)(1.033,0.482){9}{\rule{0.900pt}{0.116pt}}
\multiput(1325.00,780.17)(10.132,6.000){2}{\rule{0.450pt}{0.400pt}}
\multiput(1337.00,787.59)(0.943,0.482){9}{\rule{0.833pt}{0.116pt}}
\multiput(1337.00,786.17)(9.270,6.000){2}{\rule{0.417pt}{0.400pt}}
\multiput(1348.00,793.59)(1.033,0.482){9}{\rule{0.900pt}{0.116pt}}
\multiput(1348.00,792.17)(10.132,6.000){2}{\rule{0.450pt}{0.400pt}}
\put(181,148){\raisebox{-.8pt}{\makebox(0,0){$\bullet$}}}
\put(378,269){\raisebox{-.8pt}{\makebox(0,0){$\bullet$}}}
\put(574,388){\raisebox{-.8pt}{\makebox(0,0){$\bullet$}}}
\put(771,499){\raisebox{-.8pt}{\makebox(0,0){$\bullet$}}}
\put(967,605){\raisebox{-.8pt}{\makebox(0,0){$\bullet$}}}
\put(1164,709){\raisebox{-.8pt}{\makebox(0,0){$\bullet$}}}
\put(1360,803){\raisebox{-.8pt}{\makebox(0,0){$\bullet$}}}
\end{picture}
\caption{ Plot of $F_{\pi}$, as a function of $e$, for which the
energy $M+E_N$ is equal to the nucleon mass 939 MeV; the pion mass parameter
$m_{\pi}$ set to its experimental value of 138 MeV. Our results (bold
circles) are compared with those obtained using the rigid body
approach taken in \cite{AN:1984} (solid line). }

\end{center}
\end{figure}

In Fig.1 and Fig.2, the results of our simulations are shown in
comparison with results obtained using the rigid body approximation
used in \cite{AN:1984}. The pion mass $m_\pi$ was set to its
experimental value of 138 MeV in Fig.1, whereas in Fig.2 it was set to
the larger value of 345MeV suggested in \cite{BKS:2005}; we see that
the nucleon deformation only becomes noticeable at high values of the
pion mass.

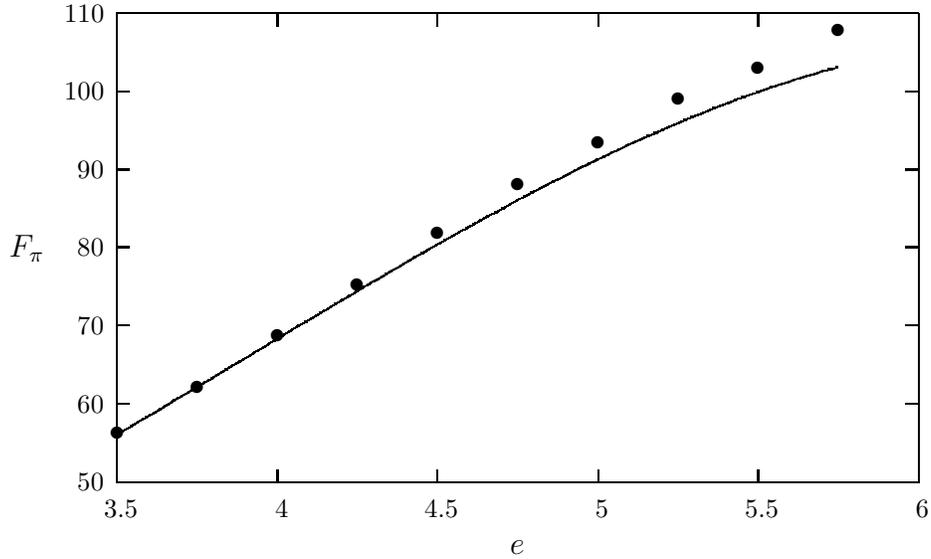
\begin{figure}
\begin{center}

% GNUPLOT: LaTeX picture
\setlength{\unitlength}{0.240900pt}
\ifx\plotpoint\undefined\newsavebox{\plotpoint}\fi
\sbox{\plotpoint}{\rule[-0.200pt]{0.400pt}{0.400pt}}%
\begin{picture}(1500,900)(0,0)
\font\gnuplot=cmr10 at 10pt
\gnuplot
\sbox{\plotpoint}{\rule[-0.200pt]{0.400pt}{0.400pt}}%
\put(181.0,123.0){\rule[-0.200pt]{4.818pt}{0.400pt}}
\put(161,123){\makebox(0,0)[r]{ 50}}
\put(1419.0,123.0){\rule[-0.200pt]{4.818pt}{0.400pt}}
\put(181.0,246.0){\rule[-0.200pt]{4.818pt}{0.400pt}}
\put(161,246){\makebox(0,0)[r]{ 60}}
\put(1419.0,246.0){\rule[-0.200pt]{4.818pt}{0.400pt}}
\put(181.0,369.0){\rule[-0.200pt]{4.818pt}{0.400pt}}
\put(161,369){\makebox(0,0)[r]{ 70}}
\put(1419.0,369.0){\rule[-0.200pt]{4.818pt}{0.400pt}}
\put(181.0,492.0){\rule[-0.200pt]{4.818pt}{0.400pt}}
\put(161,492){\makebox(0,0)[r]{ 80}}
\put(1419.0,492.0){\rule[-0.200pt]{4.818pt}{0.400pt}}
\put(181.0,614.0){\rule[-0.200pt]{4.818pt}{0.400pt}}
\put(161,614){\makebox(0,0)[r]{ 90}}
\put(1419.0,614.0){\rule[-0.200pt]{4.818pt}{0.400pt}}
\put(181.0,737.0){\rule[-0.200pt]{4.818pt}{0.400pt}}
\put(161,737){\makebox(0,0)[r]{ 100}}
\put(1419.0,737.0){\rule[-0.200pt]{4.818pt}{0.400pt}}
\put(181.0,860.0){\rule[-0.200pt]{4.818pt}{0.400pt}}
\put(161,860){\makebox(0,0)[r]{ 110}}
\put(1419.0,860.0){\rule[-0.200pt]{4.818pt}{0.400pt}}
\put(181.0,123.0){\rule[-0.200pt]{0.400pt}{4.818pt}}
\put(181,82){\makebox(0,0){ 3.5}}
\put(181.0,840.0){\rule[-0.200pt]{0.400pt}{4.818pt}}
\put(433.0,123.0){\rule[-0.200pt]{0.400pt}{4.818pt}}
\put(433,82){\makebox(0,0){ 4}}
\put(433.0,840.0){\rule[-0.200pt]{0.400pt}{4.818pt}}
\put(684.0,123.0){\rule[-0.200pt]{0.400pt}{4.818pt}}
\put(684,82){\makebox(0,0){ 4.5}}
\put(684.0,840.0){\rule[-0.200pt]{0.400pt}{4.818pt}}
\put(936.0,123.0){\rule[-0.200pt]{0.400pt}{4.818pt}}
\put(936,82){\makebox(0,0){ 5}}
\put(936.0,840.0){\rule[-0.200pt]{0.400pt}{4.818pt}}
\put(1187.0,123.0){\rule[-0.200pt]{0.400pt}{4.818pt}}
\put(1187,82){\makebox(0,0){ 5.5}}
\put(1187.0,840.0){\rule[-0.200pt]{0.400pt}{4.818pt}}
\put(1439.0,123.0){\rule[-0.200pt]{0.400pt}{4.818pt}}
\put(1439,82){\makebox(0,0){ 6}}
\put(1439.0,840.0){\rule[-0.200pt]{0.400pt}{4.818pt}}
\put(181.0,123.0){\rule[-0.200pt]{303.052pt}{0.400pt}}
\put(1439.0,123.0){\rule[-0.200pt]{0.400pt}{177.543pt}}
\put(181.0,860.0){\rule[-0.200pt]{303.052pt}{0.400pt}}
\put(40,491){\makebox(0,0){$F_{\pi}$}}
\put(810,21){\makebox(0,0){$e$}}
\put(181.0,123.0){\rule[-0.200pt]{0.400pt}{177.543pt}}
\put(181,198){\usebox{\plotpoint}}
\multiput(181.00,198.59)(0.943,0.482){9}{\rule{0.833pt}{0.116pt}}
\multiput(181.00,197.17)(9.270,6.000){2}{\rule{0.417pt}{0.400pt}}
\multiput(192.00,204.59)(0.874,0.485){11}{\rule{0.786pt}{0.117pt}}
\multiput(192.00,203.17)(10.369,7.000){2}{\rule{0.393pt}{0.400pt}}
\multiput(204.00,211.59)(0.798,0.485){11}{\rule{0.729pt}{0.117pt}}
\multiput(204.00,210.17)(9.488,7.000){2}{\rule{0.364pt}{0.400pt}}
\multiput(215.00,218.59)(0.874,0.485){11}{\rule{0.786pt}{0.117pt}}
\multiput(215.00,217.17)(10.369,7.000){2}{\rule{0.393pt}{0.400pt}}
\multiput(227.00,225.59)(0.943,0.482){9}{\rule{0.833pt}{0.116pt}}
\multiput(227.00,224.17)(9.270,6.000){2}{\rule{0.417pt}{0.400pt}}
\multiput(238.00,231.59)(0.874,0.485){11}{\rule{0.786pt}{0.117pt}}
\multiput(238.00,230.17)(10.369,7.000){2}{\rule{0.393pt}{0.400pt}}
\multiput(250.00,238.59)(0.798,0.485){11}{\rule{0.729pt}{0.117pt}}
\multiput(250.00,237.17)(9.488,7.000){2}{\rule{0.364pt}{0.400pt}}
\multiput(261.00,245.59)(0.798,0.485){11}{\rule{0.729pt}{0.117pt}}
\multiput(261.00,244.17)(9.488,7.000){2}{\rule{0.364pt}{0.400pt}}
\multiput(272.00,252.59)(0.874,0.485){11}{\rule{0.786pt}{0.117pt}}
\multiput(272.00,251.17)(10.369,7.000){2}{\rule{0.393pt}{0.400pt}}
\multiput(284.00,259.59)(0.943,0.482){9}{\rule{0.833pt}{0.116pt}}
\multiput(284.00,258.17)(9.270,6.000){2}{\rule{0.417pt}{0.400pt}}
\multiput(295.00,265.59)(0.874,0.485){11}{\rule{0.786pt}{0.117pt}}
\multiput(295.00,264.17)(10.369,7.000){2}{\rule{0.393pt}{0.400pt}}
\multiput(307.00,272.59)(0.798,0.485){11}{\rule{0.729pt}{0.117pt}}
\multiput(307.00,271.17)(9.488,7.000){2}{\rule{0.364pt}{0.400pt}}
\multiput(318.00,279.59)(0.874,0.485){11}{\rule{0.786pt}{0.117pt}}
\multiput(318.00,278.17)(10.369,7.000){2}{\rule{0.393pt}{0.400pt}}
\multiput(330.00,286.59)(0.798,0.485){11}{\rule{0.729pt}{0.117pt}}
\multiput(330.00,285.17)(9.488,7.000){2}{\rule{0.364pt}{0.400pt}}
\multiput(341.00,293.59)(0.874,0.485){11}{\rule{0.786pt}{0.117pt}}
\multiput(341.00,292.17)(10.369,7.000){2}{\rule{0.393pt}{0.400pt}}
\multiput(353.00,300.59)(0.798,0.485){11}{\rule{0.729pt}{0.117pt}}
\multiput(353.00,299.17)(9.488,7.000){2}{\rule{0.364pt}{0.400pt}}
\multiput(364.00,307.59)(0.943,0.482){9}{\rule{0.833pt}{0.116pt}}
\multiput(364.00,306.17)(9.270,6.000){2}{\rule{0.417pt}{0.400pt}}
\multiput(375.00,313.59)(0.874,0.485){11}{\rule{0.786pt}{0.117pt}}
\multiput(375.00,312.17)(10.369,7.000){2}{\rule{0.393pt}{0.400pt}}
\multiput(387.00,320.59)(0.798,0.485){11}{\rule{0.729pt}{0.117pt}}
\multiput(387.00,319.17)(9.488,7.000){2}{\rule{0.364pt}{0.400pt}}
\multiput(398.00,327.59)(0.874,0.485){11}{\rule{0.786pt}{0.117pt}}
\multiput(398.00,326.17)(10.369,7.000){2}{\rule{0.393pt}{0.400pt}}
\multiput(410.00,334.59)(0.798,0.485){11}{\rule{0.729pt}{0.117pt}}
\multiput(410.00,333.17)(9.488,7.000){2}{\rule{0.364pt}{0.400pt}}
\multiput(421.00,341.59)(0.874,0.485){11}{\rule{0.786pt}{0.117pt}}
\multiput(421.00,340.17)(10.369,7.000){2}{\rule{0.393pt}{0.400pt}}
\multiput(433.00,348.59)(0.798,0.485){11}{\rule{0.729pt}{0.117pt}}
\multiput(433.00,347.17)(9.488,7.000){2}{\rule{0.364pt}{0.400pt}}
\multiput(444.00,355.59)(0.798,0.485){11}{\rule{0.729pt}{0.117pt}}
\multiput(444.00,354.17)(9.488,7.000){2}{\rule{0.364pt}{0.400pt}}
\multiput(455.00,362.59)(0.874,0.485){11}{\rule{0.786pt}{0.117pt}}
\multiput(455.00,361.17)(10.369,7.000){2}{\rule{0.393pt}{0.400pt}}
\multiput(467.00,369.59)(0.943,0.482){9}{\rule{0.833pt}{0.116pt}}
\multiput(467.00,368.17)(9.270,6.000){2}{\rule{0.417pt}{0.400pt}}
\multiput(478.00,375.59)(0.874,0.485){11}{\rule{0.786pt}{0.117pt}}
\multiput(478.00,374.17)(10.369,7.000){2}{\rule{0.393pt}{0.400pt}}
\multiput(490.00,382.59)(0.798,0.485){11}{\rule{0.729pt}{0.117pt}}
\multiput(490.00,381.17)(9.488,7.000){2}{\rule{0.364pt}{0.400pt}}
\multiput(501.00,389.59)(0.874,0.485){11}{\rule{0.786pt}{0.117pt}}
\multiput(501.00,388.17)(10.369,7.000){2}{\rule{0.393pt}{0.400pt}}
\multiput(513.00,396.59)(0.798,0.485){11}{\rule{0.729pt}{0.117pt}}
\multiput(513.00,395.17)(9.488,7.000){2}{\rule{0.364pt}{0.400pt}}
\multiput(524.00,403.59)(0.874,0.485){11}{\rule{0.786pt}{0.117pt}}
\multiput(524.00,402.17)(10.369,7.000){2}{\rule{0.393pt}{0.400pt}}
\multiput(536.00,410.59)(0.943,0.482){9}{\rule{0.833pt}{0.116pt}}
\multiput(536.00,409.17)(9.270,6.000){2}{\rule{0.417pt}{0.400pt}}
\multiput(547.00,416.59)(0.798,0.485){11}{\rule{0.729pt}{0.117pt}}
\multiput(547.00,415.17)(9.488,7.000){2}{\rule{0.364pt}{0.400pt}}
\multiput(558.00,423.59)(0.874,0.485){11}{\rule{0.786pt}{0.117pt}}
\multiput(558.00,422.17)(10.369,7.000){2}{\rule{0.393pt}{0.400pt}}
\multiput(570.00,430.59)(0.798,0.485){11}{\rule{0.729pt}{0.117pt}}
\multiput(570.00,429.17)(9.488,7.000){2}{\rule{0.364pt}{0.400pt}}
\multiput(581.00,437.59)(1.033,0.482){9}{\rule{0.900pt}{0.116pt}}
\multiput(581.00,436.17)(10.132,6.000){2}{\rule{0.450pt}{0.400pt}}
\multiput(593.00,443.59)(0.798,0.485){11}{\rule{0.729pt}{0.117pt}}
\multiput(593.00,442.17)(9.488,7.000){2}{\rule{0.364pt}{0.400pt}}
\multiput(604.00,450.59)(0.874,0.485){11}{\rule{0.786pt}{0.117pt}}
\multiput(604.00,449.17)(10.369,7.000){2}{\rule{0.393pt}{0.400pt}}
\multiput(616.00,457.59)(0.943,0.482){9}{\rule{0.833pt}{0.116pt}}
\multiput(616.00,456.17)(9.270,6.000){2}{\rule{0.417pt}{0.400pt}}
\multiput(627.00,463.59)(0.798,0.485){11}{\rule{0.729pt}{0.117pt}}
\multiput(627.00,462.17)(9.488,7.000){2}{\rule{0.364pt}{0.400pt}}
\multiput(638.00,470.59)(0.874,0.485){11}{\rule{0.786pt}{0.117pt}}
\multiput(638.00,469.17)(10.369,7.000){2}{\rule{0.393pt}{0.400pt}}
\multiput(650.00,477.59)(0.943,0.482){9}{\rule{0.833pt}{0.116pt}}
\multiput(650.00,476.17)(9.270,6.000){2}{\rule{0.417pt}{0.400pt}}
\multiput(661.00,483.59)(0.874,0.485){11}{\rule{0.786pt}{0.117pt}}
\multiput(661.00,482.17)(10.369,7.000){2}{\rule{0.393pt}{0.400pt}}
\multiput(673.00,490.59)(0.943,0.482){9}{\rule{0.833pt}{0.116pt}}
\multiput(673.00,489.17)(9.270,6.000){2}{\rule{0.417pt}{0.400pt}}
\multiput(684.00,496.59)(0.874,0.485){11}{\rule{0.786pt}{0.117pt}}
\multiput(684.00,495.17)(10.369,7.000){2}{\rule{0.393pt}{0.400pt}}
\multiput(696.00,503.59)(0.943,0.482){9}{\rule{0.833pt}{0.116pt}}
\multiput(696.00,502.17)(9.270,6.000){2}{\rule{0.417pt}{0.400pt}}
\multiput(707.00,509.59)(0.874,0.485){11}{\rule{0.786pt}{0.117pt}}
\multiput(707.00,508.17)(10.369,7.000){2}{\rule{0.393pt}{0.400pt}}
\multiput(719.00,516.59)(0.943,0.482){9}{\rule{0.833pt}{0.116pt}}
\multiput(719.00,515.17)(9.270,6.000){2}{\rule{0.417pt}{0.400pt}}
\multiput(730.00,522.59)(0.943,0.482){9}{\rule{0.833pt}{0.116pt}}
\multiput(730.00,521.17)(9.270,6.000){2}{\rule{0.417pt}{0.400pt}}
\multiput(741.00,528.59)(0.874,0.485){11}{\rule{0.786pt}{0.117pt}}
\multiput(741.00,527.17)(10.369,7.000){2}{\rule{0.393pt}{0.400pt}}
\multiput(753.00,535.59)(0.943,0.482){9}{\rule{0.833pt}{0.116pt}}
\multiput(753.00,534.17)(9.270,6.000){2}{\rule{0.417pt}{0.400pt}}
\multiput(764.00,541.59)(1.033,0.482){9}{\rule{0.900pt}{0.116pt}}
\multiput(764.00,540.17)(10.132,6.000){2}{\rule{0.450pt}{0.400pt}}
\multiput(776.00,547.59)(0.943,0.482){9}{\rule{0.833pt}{0.116pt}}
\multiput(776.00,546.17)(9.270,6.000){2}{\rule{0.417pt}{0.400pt}}
\multiput(787.00,553.59)(0.874,0.485){11}{\rule{0.786pt}{0.117pt}}
\multiput(787.00,552.17)(10.369,7.000){2}{\rule{0.393pt}{0.400pt}}
\multiput(799.00,560.59)(0.943,0.482){9}{\rule{0.833pt}{0.116pt}}
\multiput(799.00,559.17)(9.270,6.000){2}{\rule{0.417pt}{0.400pt}}
\multiput(810.00,566.59)(0.943,0.482){9}{\rule{0.833pt}{0.116pt}}
\multiput(810.00,565.17)(9.270,6.000){2}{\rule{0.417pt}{0.400pt}}
\multiput(821.00,572.59)(1.033,0.482){9}{\rule{0.900pt}{0.116pt}}
\multiput(821.00,571.17)(10.132,6.000){2}{\rule{0.450pt}{0.400pt}}
\multiput(833.00,578.59)(0.943,0.482){9}{\rule{0.833pt}{0.116pt}}
\multiput(833.00,577.17)(9.270,6.000){2}{\rule{0.417pt}{0.400pt}}
\multiput(844.00,584.59)(1.033,0.482){9}{\rule{0.900pt}{0.116pt}}
\multiput(844.00,583.17)(10.132,6.000){2}{\rule{0.450pt}{0.400pt}}
\multiput(856.00,590.59)(0.943,0.482){9}{\rule{0.833pt}{0.116pt}}
\multiput(856.00,589.17)(9.270,6.000){2}{\rule{0.417pt}{0.400pt}}
\multiput(867.00,596.59)(1.267,0.477){7}{\rule{1.060pt}{0.115pt}}
\multiput(867.00,595.17)(9.800,5.000){2}{\rule{0.530pt}{0.400pt}}
\multiput(879.00,601.59)(0.943,0.482){9}{\rule{0.833pt}{0.116pt}}
\multiput(879.00,600.17)(9.270,6.000){2}{\rule{0.417pt}{0.400pt}}
\multiput(890.00,607.59)(0.943,0.482){9}{\rule{0.833pt}{0.116pt}}
\multiput(890.00,606.17)(9.270,6.000){2}{\rule{0.417pt}{0.400pt}}
\multiput(901.00,613.59)(1.033,0.482){9}{\rule{0.900pt}{0.116pt}}
\multiput(901.00,612.17)(10.132,6.000){2}{\rule{0.450pt}{0.400pt}}
\multiput(913.00,619.59)(1.155,0.477){7}{\rule{0.980pt}{0.115pt}}
\multiput(913.00,618.17)(8.966,5.000){2}{\rule{0.490pt}{0.400pt}}
\multiput(924.00,624.59)(1.033,0.482){9}{\rule{0.900pt}{0.116pt}}
\multiput(924.00,623.17)(10.132,6.000){2}{\rule{0.450pt}{0.400pt}}
\multiput(936.00,630.59)(1.155,0.477){7}{\rule{0.980pt}{0.115pt}}
\multiput(936.00,629.17)(8.966,5.000){2}{\rule{0.490pt}{0.400pt}}
\multiput(947.00,635.59)(1.033,0.482){9}{\rule{0.900pt}{0.116pt}}
\multiput(947.00,634.17)(10.132,6.000){2}{\rule{0.450pt}{0.400pt}}
\multiput(959.00,641.59)(1.155,0.477){7}{\rule{0.980pt}{0.115pt}}
\multiput(959.00,640.17)(8.966,5.000){2}{\rule{0.490pt}{0.400pt}}
\multiput(970.00,646.59)(1.033,0.482){9}{\rule{0.900pt}{0.116pt}}
\multiput(970.00,645.17)(10.132,6.000){2}{\rule{0.450pt}{0.400pt}}
\multiput(982.00,652.59)(1.155,0.477){7}{\rule{0.980pt}{0.115pt}}
\multiput(982.00,651.17)(8.966,5.000){2}{\rule{0.490pt}{0.400pt}}
\multiput(993.00,657.59)(1.155,0.477){7}{\rule{0.980pt}{0.115pt}}
\multiput(993.00,656.17)(8.966,5.000){2}{\rule{0.490pt}{0.400pt}}
\multiput(1004.00,662.59)(1.267,0.477){7}{\rule{1.060pt}{0.115pt}}
\multiput(1004.00,661.17)(9.800,5.000){2}{\rule{0.530pt}{0.400pt}}
\multiput(1016.00,667.59)(1.155,0.477){7}{\rule{0.980pt}{0.115pt}}
\multiput(1016.00,666.17)(8.966,5.000){2}{\rule{0.490pt}{0.400pt}}
\multiput(1027.00,672.59)(1.267,0.477){7}{\rule{1.060pt}{0.115pt}}
\multiput(1027.00,671.17)(9.800,5.000){2}{\rule{0.530pt}{0.400pt}}
\multiput(1039.00,677.59)(1.155,0.477){7}{\rule{0.980pt}{0.115pt}}
\multiput(1039.00,676.17)(8.966,5.000){2}{\rule{0.490pt}{0.400pt}}
\multiput(1050.00,682.59)(1.267,0.477){7}{\rule{1.060pt}{0.115pt}}
\multiput(1050.00,681.17)(9.800,5.000){2}{\rule{0.530pt}{0.400pt}}
\multiput(1062.00,687.59)(1.155,0.477){7}{\rule{0.980pt}{0.115pt}}
\multiput(1062.00,686.17)(8.966,5.000){2}{\rule{0.490pt}{0.400pt}}
\multiput(1073.00,692.59)(1.155,0.477){7}{\rule{0.980pt}{0.115pt}}
\multiput(1073.00,691.17)(8.966,5.000){2}{\rule{0.490pt}{0.400pt}}
\multiput(1084.00,697.60)(1.651,0.468){5}{\rule{1.300pt}{0.113pt}}
\multiput(1084.00,696.17)(9.302,4.000){2}{\rule{0.650pt}{0.400pt}}
\multiput(1096.00,701.59)(1.155,0.477){7}{\rule{0.980pt}{0.115pt}}
\multiput(1096.00,700.17)(8.966,5.000){2}{\rule{0.490pt}{0.400pt}}
\multiput(1107.00,706.59)(1.267,0.477){7}{\rule{1.060pt}{0.115pt}}
\multiput(1107.00,705.17)(9.800,5.000){2}{\rule{0.530pt}{0.400pt}}
\multiput(1119.00,711.60)(1.505,0.468){5}{\rule{1.200pt}{0.113pt}}
\multiput(1119.00,710.17)(8.509,4.000){2}{\rule{0.600pt}{0.400pt}}
\multiput(1130.00,715.60)(1.651,0.468){5}{\rule{1.300pt}{0.113pt}}
\multiput(1130.00,714.17)(9.302,4.000){2}{\rule{0.650pt}{0.400pt}}
\multiput(1142.00,719.59)(1.155,0.477){7}{\rule{0.980pt}{0.115pt}}
\multiput(1142.00,718.17)(8.966,5.000){2}{\rule{0.490pt}{0.400pt}}
\multiput(1153.00,724.60)(1.651,0.468){5}{\rule{1.300pt}{0.113pt}}
\multiput(1153.00,723.17)(9.302,4.000){2}{\rule{0.650pt}{0.400pt}}
\multiput(1165.00,728.60)(1.505,0.468){5}{\rule{1.200pt}{0.113pt}}
\multiput(1165.00,727.17)(8.509,4.000){2}{\rule{0.600pt}{0.400pt}}
\multiput(1176.00,732.60)(1.505,0.468){5}{\rule{1.200pt}{0.113pt}}
\multiput(1176.00,731.17)(8.509,4.000){2}{\rule{0.600pt}{0.400pt}}
\multiput(1187.00,736.60)(1.651,0.468){5}{\rule{1.300pt}{0.113pt}}
\multiput(1187.00,735.17)(9.302,4.000){2}{\rule{0.650pt}{0.400pt}}
\multiput(1199.00,740.60)(1.505,0.468){5}{\rule{1.200pt}{0.113pt}}
\multiput(1199.00,739.17)(8.509,4.000){2}{\rule{0.600pt}{0.400pt}}
\multiput(1210.00,744.60)(1.651,0.468){5}{\rule{1.300pt}{0.113pt}}
\multiput(1210.00,743.17)(9.302,4.000){2}{\rule{0.650pt}{0.400pt}}
\multiput(1222.00,748.61)(2.248,0.447){3}{\rule{1.567pt}{0.108pt}}
\multiput(1222.00,747.17)(7.748,3.000){2}{\rule{0.783pt}{0.400pt}}
\multiput(1233.00,751.60)(1.651,0.468){5}{\rule{1.300pt}{0.113pt}}
\multiput(1233.00,750.17)(9.302,4.000){2}{\rule{0.650pt}{0.400pt}}
\multiput(1245.00,755.60)(1.505,0.468){5}{\rule{1.200pt}{0.113pt}}
\multiput(1245.00,754.17)(8.509,4.000){2}{\rule{0.600pt}{0.400pt}}
\multiput(1256.00,759.61)(2.248,0.447){3}{\rule{1.567pt}{0.108pt}}
\multiput(1256.00,758.17)(7.748,3.000){2}{\rule{0.783pt}{0.400pt}}
\multiput(1267.00,762.61)(2.472,0.447){3}{\rule{1.700pt}{0.108pt}}
\multiput(1267.00,761.17)(8.472,3.000){2}{\rule{0.850pt}{0.400pt}}
\multiput(1279.00,765.60)(1.505,0.468){5}{\rule{1.200pt}{0.113pt}}
\multiput(1279.00,764.17)(8.509,4.000){2}{\rule{0.600pt}{0.400pt}}
\multiput(1290.00,769.61)(2.472,0.447){3}{\rule{1.700pt}{0.108pt}}
\multiput(1290.00,768.17)(8.472,3.000){2}{\rule{0.850pt}{0.400pt}}
\multiput(1302.00,772.61)(2.248,0.447){3}{\rule{1.567pt}{0.108pt}}
\multiput(1302.00,771.17)(7.748,3.000){2}{\rule{0.783pt}{0.400pt}}
\put(181,203){\raisebox{-.8pt}{\makebox(0,0){$\bullet$}}}
\put(307,275){\raisebox{-.8pt}{\makebox(0,0){$\bullet$}}}
\put(433,356){\raisebox{-.8pt}{\makebox(0,0){$\bullet$}}}
\put(558,436){\raisebox{-.8pt}{\makebox(0,0){$\bullet$}}}
\put(684,516){\raisebox{-.8pt}{\makebox(0,0){$\bullet$}}}
\put(810,594){\raisebox{-.8pt}{\makebox(0,0){$\bullet$}}}
\put(936,659){\raisebox{-.8pt}{\makebox(0,0){$\bullet$}}}
\put(1062,727){\raisebox{-.8pt}{\makebox(0,0){$\bullet$}}}
\put(1187,776){\raisebox{-.8pt}{\makebox(0,0){$\bullet$}}}
\put(1313,835){\raisebox{-.8pt}{\makebox(0,0){$\bullet$}}}
\end{picture}
\end{center}
\caption{ As Fig. 1, but with the pion mass parameter $m_{\pi}$ set to the value of 345 MeV suggested in \cite{BKS:2005}. Our results (bold circles) are compared with those obtained using the rigid body approach taken in \cite{AN:1984} (solid line). } 
\end{figure}

\section{Discussion}

Although it has appealing mathematical and physical properties, the
Skyrme model has only been modestly successful in modeling nuclear
dynamics. Perhaps one of the biggest challenges is that the zero-mode
quantization of the classical minimum fails to give even the correct
lowest energy state for many values of the baryon number
\cite{I:2000,S:2003}. It is possible that this is because
inappropriate parameter values are used to calculate the classical
minima: for example, in \cite{BS:2005} it is suggested that using a
non-zero, or even an unphysically large, pion mass will significantly
affect the structure of higher charge Skyrmions and alter the symmetries on which the quantization is based \cite{S:2005}. Another possibility
is that the classical minimum is inappropriate and that the energy
minimum of the effective quantum Hamiltonian should be used for
zero-mode quantization. In this paper this has been done in the simple
case of a axial symmetric nucleon; the higher charge cases will be
more challenging computationally but should follow in a similar way.

\section*{Acknowledgments}
SM acknowledges receipt of funding under the Programme for Research in
Third-Level Institutions (PRTLI), administered by the HEA; CJH and SM
acknowledge a Trinity College Dublin start-up grant. We also thank the
Trinity Centre for High-Performance Computing for the use of their
computing facilities.

\section*{Appendix: Identification of normal moments of inertia}
Following \cite{BC:1988}, we establish a relation between $U_3, V_3$
and $W_3$ for axially symmetric fields. We first express the quantity $\epsilon_{3jk}x_j R_k$ in polar coordinates:
\begin{eqnarray}
 \epsilon_{3jk}x_j R_k 
&=& \left( \mathbf{x} \times \nabla \right)_3 U U^{\dagger} 
\nonumber \\[5pt]
&=&\frac{\partial U}{\partial \phi}U^{\dagger}.
\end{eqnarray}
We can then identify $\epsilon_{ij3}x_j R_i$ and $- \frac{i}{2}\left[ \sigma_3, U \right]U^{\dagger}$ by looking at the general form
of an axially symmetric field with unit baryon number:
\begin{equation}
U= \mbox{e}^{\frac{-i \sigma_3 \phi}{2}}\mbox{e}^{i f \left( r,z \right)n_i \sigma_i }
\mbox{e}^{\frac{i \sigma_3 \phi}{2}}
\end{equation}
and taking its derivative with respect to $\phi$:
\begin{eqnarray}
\frac{\partial U}{\partial \phi}
&=& -{\frac{i \sigma_3}{2}}\left( \mbox{e}^{- \frac{i \sigma_3 \phi}{2}}\mbox{e}^{i f 
\left( r,z \right)}
\mbox{e}^{\frac{i \sigma_3 \phi}{2}}\right)
+ \left(\mbox{e}^{\frac{- i \sigma_3 \phi}{2}}\mbox{e}^{i f \left( r,z \right)}
\mbox{e}^{\frac{i \sigma_3 \phi}{2}}\right)
{\frac{i\sigma_3}{2}}
\nonumber \\[5pt]
&=& {-\frac{i}{2}}\left[ \sigma_3, U \right].
\end{eqnarray}
We see the expressions for the inertias $U_{3}, V_{3}$ and $W_{3}$ in the moment of inertia integrals (\ref{clinertia})
differ only in the terms $\epsilon_{3jk}x_j R_k$ and
$- \frac{i}{2}\left[ \sigma_3, U \right]U^{\dagger}$, and so
\begin{equation}
U_3=V_3=W_3.
\end{equation}

\end{document}